\documentclass[aps,prd,twocolumn,groupedaddress]{revtex4-2}
\usepackage{graphicx}
\usepackage{amsmath}
\usepackage{dcolumn}
\usepackage{epsfig}
\usepackage{psfrag}
\usepackage{subfigure}
\usepackage{hyperref}

\RequirePackage{xspace}

\def\TeV{\ifmmode {\mathrm{\ Te\kern -0.1em V}}\else
	                   \textrm{Te\kern -0.1em V}\fi}%
\def\GeV{\ifmmode {\mathrm{\ Ge\kern -0.1em V}}\else
	                   \textrm{Ge\kern -0.1em V}\fi}%
\def\MeV{\ifmmode {\mathrm{\ Me\kern -0.1em V}}\else
	                   \textrm{Me\kern -0.1em V}\fi}%
\def\keV{\ifmmode {\mathrm{\ ke\kern -0.1em V}}\else
	                   \textrm{ke\kern -0.1em V}\fi}%
\def\eV{\ifmmode  {\mathrm{\ e\kern -0.1em V}}\else
	                   \textrm{e\kern -0.1em V}\fi}%
\let\tev=\TeV
\let\gev=\GeV

\def\TeVc{\ifmmode {\mathrm{\ Te\kern -0.1em V}/c}\else
	                   {\textrm{Te\kern -0.1em V}/$c$}\fi}%
\def\GeVc{\ifmmode {\mathrm{\ Ge\kern -0.1em V}/c}\else
	                   {\textrm{Ge\kern -0.1em V}/$c$}\fi}%
\def\MeVc{\ifmmode {\mathrm{\ Me\kern -0.1em V}/c}\else
	                   {\textrm{Me\kern -0.1em V}/$c$}\fi}%
\def\keVc{\ifmmode {\mathrm{\ ke\kern -0.1em V}/c}\else
	                   {\textrm{ke\kern -0.1em V}/$c$}\fi}%
\def\eVc{\ifmmode  {\mathrm{\ e\kern -0.1em V}/c}\else
	                   {\textrm{e\kern -0.1em V}/$c$}\fi}%
	
\def\cm{\ifmmode  {\mathrm{\ cm}}\else	                   
\textrm{~cm}\fi}%

\def\mm{\ifmmode  {\mathrm{\ mm}}\else	                   
\textrm{~mm}\fi}%
	
\usepackage{relsize}
\def\babar{\mbox{\slshape B\kern-0.1em{\smaller A}\kern-0.1em
    B\kern-0.1em{\smaller A\kern-0.2em R}}}

\begin{document}

\title{Reconstruction and identification of $H\to WW^*$ with high transverse momentum\\ in the full hadronic final state }

\author{Chunhui Chen} 
\affiliation{Department of Physics and Astronomy, Iowa State University, Ames, Iowa 50011, USA}

\begin{abstract}
This paper presents a study of the reconstruction and identification of $H\to WW^*$ with high transverse momentum,
 where both $W^{(*)}$ bosons decay hadronically. We show that the boosted $H\to WW^*$ can be effectively reconstructed as a single jet and 
 identified using jet substructures in the center-of-mass frame of the jet. 
Such a  reconstruction and identification approach can discriminate the boosted 
$H\to WW^*$ in the full hadronic final state from QCD jets.  This result will significantly improve 
experimental sensitivities of searches for potential new physics phenomena beyond the standard model in 
final states containing highly boosted Higgs bosons.
\end{abstract}

\maketitle

\section{Introduction}

Many new physics (NP) extensions to the standard model (SM) predict new particles with masses at the TeV scale. 
Some of these heavy resonances~\cite{Schmaltz:2005ky,Agashe:2007ki,Agashe:2008jb,Branco:2011iw,Dobrescu:2017sue} 
can decay into final states containing Higgs bosons.  The most effective way to search for such particles is to reconstruct 
the Higgs boson in its dominant decay into a bottom quark-antiquark pair ($b\bar{b}$) final state. 
Because the Higgs bosons decayed from heavy resonances have very large momenta (boosted), 
the hadronically decaying products of $H\to b\bar{b}$ are so collimated that they are often reconstructed as single jets in the experiments.
The searches for new heavy resonances using the signature of boosted $H\to b\bar{b}$ is an emerging field of research that has attracted
significant theoretical and experimental interests~\cite{Butterworth:2008iy,Abdesselam:2010pt,DiMicco:2019ngk}.
While the current searches by both the ATLAS~\cite{Aad:2014yja,Aad:2015yza,Aad:2015uka, Aad:2015xja,Aaboud:2016lwx,Aaboud:2017ecz, Aaboud:2017ahz, Aaboud:2017cxo, Aaboud:2018knk, Aaboud:2018bun, Aaboud:2018fgi, Aaboud:2018zhh,Aad:2020tps,Aad:2020ylk,Aad:2020ldt}  
and CMS~\cite{Khachatryan:2015ywa,Khachatryan:2015yea,Khachatryan:2015bma,Khachatryan:2016yji,Khachatryan:2016cfa,Khachatryan:2016cfx,
Sirunyan:2017wto,Sirunyan:2017isc,Sirunyan:2018qob,Sirunyan:2018fuh,Sirunyan:2018qca,CMS-Hgamma,Sirunyan:2019quj}  
experiments at the Large Hadron Collider (LHC) start to probe heavy resonances with masses as large as 2--3\,\tev, 
the experimental sensitivities  become limited by the signal reconstruction efficiency as the production cross sections of 
both heavy resonances and  background from the SM processes drop significantly at the very high energy scale. 

In this paper, we present a study of the reconstruction and identification of boosted $H\to WW^*$, 
 where both the on-shell $W$ boson and the off-shell $W^*$ boson decay hadronically. 
 The $H\to WW^*$ decay mode has the second-largest decaying branch fraction of $21\,\%$, and
there haven't been  any published studies on its  boosted signature. We show that a boosted $H\to WW^*$ in the hadronic final 
 state can be effectively reconstructed as a single jet, hereafter referred to as a $H\to W W^*$ jet.
 By using jet substructures in the center-of-mass  frame of the jet, the $H\to W W^*$ jets can 
 be distinguished  from QCD jets, where the QCD jets are defined as those jets initiated by a 
 non-top quark or a gluon.  This result can significantly improve 
experimental sensitivities of searches for potential NP phenomena  in final states containing highly boosted Higgs bosons.

\section{Event Sample}
We use the boosted $H\to WW^*$ jets, from the SM process of  $ZH$  production,  as a benchmark to study the
signal reconstruction and identification.  The SM multijet production process is used to model the QCD jets.
The sample of $H\to WW^*$ jets is  further reweighted on a jet-by-jet basis such that  its jet kinematics in transverse momentum ($p_T$) and 
pseudorapidity ($\eta$) match that of the QCD jet sample. 

All the events used in this analysis are produced using the P{\footnotesize ythia} 8.244 Monte Carlo (MC) 
event generator~\cite{Sjostrand:2006za,Sjostrand:2007gs} for the $pp$ collision at $13\,\rm TeV$ center-of-mass energy.
The spread of beam interaction point is assumed to follow a Gaussian distribution with a width of $0.015\,\mm$ in the transverse beam direction
and $45\,\mm$ in the longitudinal beam direction~\cite{Aad:2008zzm}.
We divide the $(\eta, \phi)$ plane into $0.1\times 0.1$ cells
to simulate the finite resolution of the calorimeter detector at the LHC experiments.
The energies of particles entering each cell in each event, except for the neutrinos, are summed over and
replaced with a massless pseudoparticle of the same energy, also referred to as an energy cluster,  pointing to the center of the  cell. 
These pseudoparticles are fed into the F{\footnotesize astJet} 3.0.1~\cite{Cacciari:2005hq} package for  jet reconstruction.
As for charged tracks, their momenta and vertex positions are smeared according to the expected resolutions of the ATLAS detector~\cite{Aad:2009wy}.
 The effect of multiple $pp$ interactions in the same event (pileup) is included by overlaying minimum-bias events simulated with
 P{\footnotesize ythia} 8.244  on each event of interest in all samples. The number of pileup events is assumed to follow a Poisson distribution 
 with a mean of 35, which is the observed average number of pileup events at ATLAS during its data taking at the $13\,\rm TeV$ $pp$ collisions.
 
 \section{Reconstruction and identification of $H\to WW^*$ jets}
 \subsection{Event selection}
Jets are reconstructed with the anti-$k_T$ algorithm~\cite{Cacciari:2008gp}  
with a distance parameter of $\Delta R=1.0$, which is the default  jet reconstruction algorithm used at the ATLAS and CMS experiments. 
 To correct the presence of additional energy depositions from underlying events and pileups, we employ a jet area correction 
 technique~\cite{Cacciari:2007fd} to take into account the effects  on an event-by-event basis.
For each event, the distribution of transverse energy densities is calculated for all jets with $|\eta|<2.1$, and its median is taken as
an estimate of the energy density of  the pileup and underlying events.  We subsequently correct each jet by subtracting the product 
of the transverse energy density and the jet area, which is determined with the ``active" area calculation technique~\cite{Cacciari:2007fd}. 
This method results in a modified jet four-momentum $p^\mu=(m_{\rm jet}, \vec{p}_{\rm jet})$ that is used throughout the paper unless explicitly stated otherwise.
 
\begin{figure}[!htb]
\begin{center}
\includegraphics[width=0.42\textwidth]{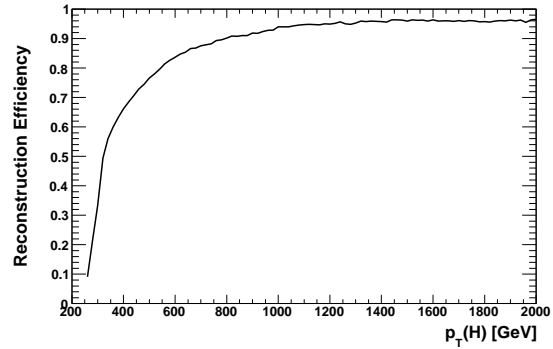}
\caption{The reconstruction efficiency of a hadronically decaying $H\to WW^*$ using a single jet as a function of the 
Higgs boson $p_T$. The Higgs bosons are required to have $|\eta|<2.0$.}
\label{fig:Reco_eff}
\end{center}
\end{figure}

 We select jets with $p_T>350\,\gev$ and $|\eta|<2.0$ as $H\to WW^*$ jet candidates.  
 MC studies show that the averaged momentum of the $W^{(*)}$ bosons decayed from the Higgs boson 
 is about 25\,\gev\ in the Higgs rest frame, much smaller than the momenta of the selected $H\to WW^*$ jet candidates.
 As a result, the decaying products of the two $W^{(*)}$ bosons often overlap with each other.
 Figure~\ref{fig:Reco_eff} shows the efficiency to reconstruct a hadronically decaying $H\to WW^*$ using a single jet 
 as a function of the Higgs boson $p_T$. The efficiency is defined as  $\epsilon=N_{\rm reco}/N_{\rm gen}$, where $N_{\rm gen}$ is the number of generated
 Higgs bosons, and $N_{\rm reco}$ is the number of reconstructed $H\to WW^*$ jet candidates whose separations between partons 
 decayed from  $W^{(*)}$ bosons and $H\to WW^*$ jet are less than 1. 
 Here the separation between two objects is defined as  $\Delta R=\sqrt{\Delta\eta^2+\Delta\phi^2}$, 
where $\Delta\eta$ and $\Delta\phi$ are  the differences in pseudorapidity and the azimuthal angle between 
the two objects' momenta, respectively.
For a Higgs boson with $p_T=350\,\gev$, the averaged separation  between the  decaying products of $H\to WW^*$ is about $0.8$, and 
approximately 55\,\% of the $H\to WW^*$ in the hadronic final state can be reconstructed as a single jet.
The averaged separation  gradually decreases as a function of $p_T$ to
be less than 0.5 for Higgs bosons with $p_T>1\,\tev$, where more than 90\,\% of the 
 $H\to WW^*$ in the hadronic final state can be reconstructed as a single jet.

\begin{figure}[!htb]
\begin{center}
\includegraphics[width=0.48\textwidth]{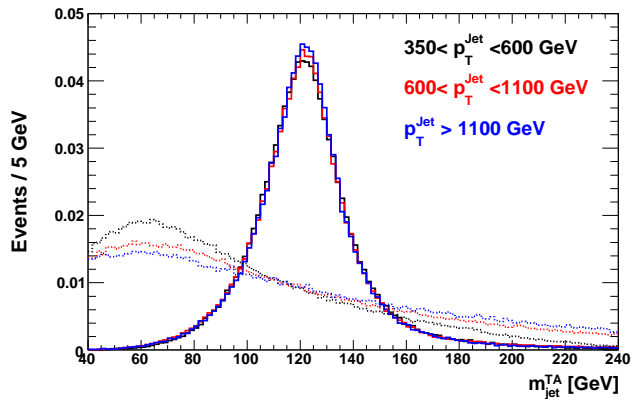}
\caption{The track-assisted jet mass distribution of signal $H\to WW^*$ jets (solid line) and QCD jets (dashed line) in different $p_T$ ranges.}
\label{fig:mJet}
\end{center}
\end{figure}

 We further require the track-assisted mass~\cite{ATLAS-CONF-2016-035} of selected $H\to WW^*$ jets to satisfy
 $40<m^{\rm TA}_{\rm jet}<240\,\gev$. The track-assisted jet mass is defined as 
 $m^{\rm TA}_{\rm jet}=m^{\rm trk}_{\rm jet}\times (p_T/p^{\rm trk}_{T,{\rm jet}})$, where $m^{\rm trk}_{\rm jet}$
 and $p_T/p^{\rm trk}_{T,{\rm jet}}$ are the invariant mass and the total transverse momentum of charged tracks associated
 with the jet, respectively. Only charged tracks with $p_{\rm T} >1\,\gev$ and $|\eta|<2.5$ are considered. They
are also required to satisfy the criteria that  $|d_0|<1\,\mm$ and $|z_0-z_{\rm pv}|\sin\theta<1.5\,\mm$,
where $d_0$ and $z_0$ are the transverse and longitudinal impact parameters of the charged track, respectively, 
$z_{\rm pv}$ is the longitudinal position of the primary vertex, and $\theta$ is the polar angle of the
charged track.  A  charged track is considered to be associated with a jet with $p_T$ in a unit of \gev\ only if the separation $\Delta R$ 
between the track and jet  is less than $R_{\rm max}$, where $R_{\rm max}=1.0-0.4\times (p_T-350)/650$ for jets with 
$p_T<1000\gev$ and $R_{\rm max}=0.6$ for jets with $p_T>1000\,\gev$, respectively. The
track-assisted jet mass distributions of signal $H\to WW^*$ jets  and QCD jets in different $p_T$ ranges
are shown in Fig.~\ref{fig:mJet}.  
The $m^{\rm TA}_{\rm jet}$ distribution of the signal jets peaks around the Higgs boson mass and its shape shows no 
significant variation as a function of jet $p_T$.

\subsection{Jet substructure}
Signal $H\to WW^*$ jets can be further distinguished from QCD jets using jet substructure variables. 
There are many jet substructure variables~\cite{Altheimer:2013yza} proposed and some of them have been successfully used to 
identify the boosted hadronically decaying $W/Z$ boson, top quark, and $H\to b\bar{b}$ boson
by the ATLAS~\cite{Aaboud:2018psm,Aad:2019uoz} and CMS~\cite{Khachatryan:2014vla,Sirunyan:2017ezt} 
experiments. As a simple illustration, this paper uses the substructure variables 
defined in the  center-of-mass frame of the jet, introduced in Ref.~\cite{Chen:2011ah}. They are
the thrust ($T$), thrust-minor ($T_{\rm min}$), sphericity ($S$), and 
the ratio between the second-order and zeroth-order Fox-Wolfram moments ($R_2$). 
Those variables  have been successfully implemented 
by the ATLAS experiment to make the first observation of the boosted hadronically decaying vector boson 
reconstructed as a single jet from the SM $W/Z$+jets production~\cite{Aad:2014haa}.
\begin{figure*}[!htb]
\begin{center}
\includegraphics[width=1\textwidth]{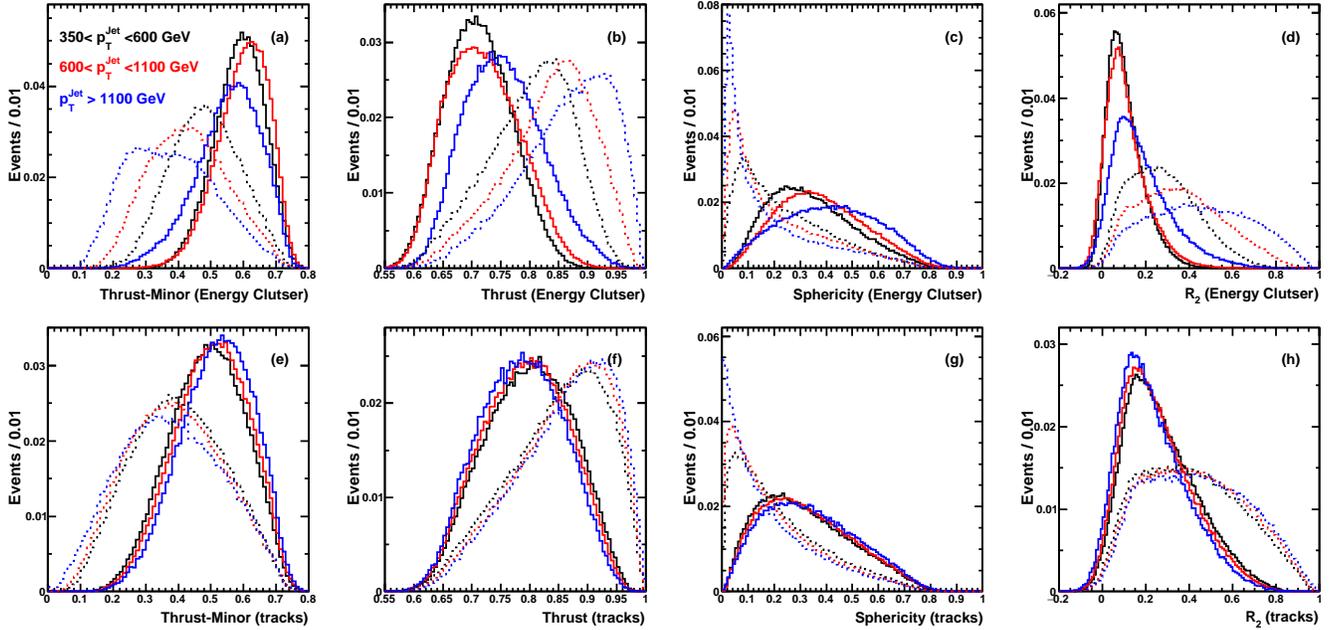}
\caption{The distributions of the jet substructure variables calculated using energy clusters (top row) of a jet or charged tracks (bottom row) 
associated with the jet for $H\to WW^*$ signal jets (solid line) and QCD jet background (dashed line) in different jet $p_T$ ranges.  
Highly directional distributions of energy clusters or charged tracks have $T\approx 1$, $T_{\rm min}\approx 0$, $S\approx 0$,
and a  large value of $R_2$, while $T\approx 0.5$, $T_{\rm min}\approx 1$, $S\approx 1$,
and a small value of $R_2$ correspond to an isotropic distribution. 
All the distributions are normalized to unity.}
\label{fig:BDT_var}
\end{center}
\end{figure*}

We define the center-of-mass frame (rest frame) of a jet as the frame where the four-momentum of the
jet is equal to $p^{\rm rest}_{\mu}\equiv (m_{\rm jet}, 0, 0, 0)$. 
All the jet substructure variables are calculated using the energy clusters of a jet or the charged tracks associated 
with a jet in the center-of-mass frame of the jet. Their distributions are shown in Fig.~\ref{fig:BDT_var}.
A jet consists of its constituent particles.
While the $H\to WW^*$ is a two-body decay, the momenta of the $W^{(*)}$ bosons in the 
Higgs rest frame are very small due to the mass suppression. As a result, the 
distribution of the constituent particles of a boosted $H\to WW^*$ in its center-of-mass frame has
a four-body decay topology, with each subjet corresponding to one quark decayed from the two $W^{(*)}$ bosons.
The hadronically decaying products from a boosted $H\to WW^*$ has a relatively 
isotropic distribution in the jet rest frame. 
It is worth noting that the tracks associated with a Higgs boson with a larger  $p_T$ are
more isotropically distributed compared to the ones associated with a Higgs boson with a smaller $p_T$,
as shown in Fig.~\ref{fig:BDT_var}. Such an effect is also seen in the distribution of jet sphericity 
calculated using the energy clusters, but not for thrust, thrust-minor, and $R_2$ 
as their calculations based on the energy clusters are degraded by the finite resolution of the Calorimeter detector when jet $p_T$ 
becomes very large.
On the other hand, the constituent particle distribution of a QCD jet in the jet rest frame is less isotropic
and becomes even more directional when jet $p_T$ becomes larger.  
As a result, the jet substructure variables defined in the jet rest frame has a better discriminating power to separate
boosted $H\to WW^*$ bosons from the background for jets with larger $p_T$.

We further recluster the energy clusters of a jet to reconstruct subjets in the jet rest frame using a generalized 
EEkT  algorithm~\cite{Catani:1991hj} in the F{\footnotesize astJet} 3.0.1~\cite{Cacciari:2005hq} package with
the parameters of  $p=0$ and $R=0.8$. 
The reconstructed subjets  are required to have energy $E_{\rm subjet}>10\,\gev$ in the jet rest frame. 
A charged track is considered to be associated with a subjet only if their angular separation is less than 0.8 in the jet rest frame. 

\subsection{Identification of $H\to WW^*$ jets}
\begin{figure}[!htb]
\begin{center}
\includegraphics[width=0.49\textwidth]{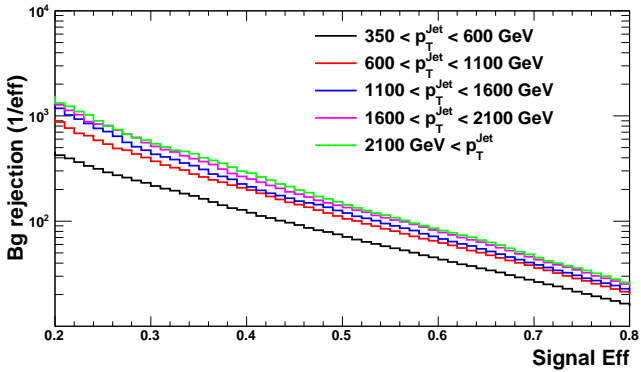}
\caption{The background rejection of QCD jets vs. the signal efficiency of $H\to WW^*$ jets in different
jet $p_T$ ranges, where the background rejection is defined as $1/\epsilon_{\rm bg}$, and $\epsilon_{\rm bg}$
is the misidentification efficiency of the background QCD jets for a given signal identification efficiency.
}
\label{fig:EffRej}
\end{center}
\end{figure}

The final variable to identify boosted $H\to WW^*$ jets  is constructed using a boosted decision tree (BDT) 
algorithm. The input variables used in the BDT include $m^{\rm TA}_{\rm jet}$,
the number of charged tracks associated with the jet, the number of subjets reconstructed in the jet rest frame,
and jet substructure  variables ($T$, $T_{\rm min}$, $S$, and $R_2$) calculated using energy clusters and charged 
tracks, respectively.
Among all the input variables in the BDT, the best one to distinguish signal from background is $m^{\rm TA}_{\rm jet}$, which also
does not have a strong correlation with the other variables. All the substructure variables have comparable discriminating powers, but they are highly 
correlated. The linear correlation efficiencies between them vary approximately between 60\,\% and 90\,\%. 
The number of subjet has slightly less discriminating power than jet substructure variables, and their correlations 
range between 20\,\% and 50\,\%. While  the number of charged tracks has the least power to disentangle signal from background, 
its correlations to other variables are less than 10\,\%. 
The signal  identification efficiency of $H\to WW^*$ jets  vs 
the background rejection of QCD jets for the BDT variable is shown in Fig~\ref{fig:EffRej} ,
where the signal identification efficiency is defined as the probability for a reconstructed signal $H\to WW^*$ jet to pass a given
requirement of the  BDT variable.
The  rejection of QCD jets for a given $H\to WW^*$ jet identification efficiency
is comparable to the $H\to b\bar{b}$ taggers at ATLAS~\cite{ATLAS:2017juw} and CMS~\cite{Sirunyan:2017ezt}. The performance 
of $H\to WW^*$ identification gets better for jets with larger $p_T$ unlike the  $H\to b\bar{b}$ taggers, whose background rejections
degrade when the jet $p_T$ increases~\cite{ATLAS:2017juw,Sirunyan:2017ezt}. 
The performance of $H\to WW^*$ identification
can be further improved by including additional BDT input variables, such as the subjet energies, the number of
charged tracks associated with each subjet, other jet substructure variables~\cite{Altheimer:2013yza} defined in the lab frame, and so on. 
Such a dedicated study is beyond the scope of this paper.

\section{Application}
Reconstructed $H\to WW^*$ jets can be used to improve searches for NP with specific final state signatures.
Here we demonstrate such an application by considering a search for a heavy resonance ($X$) with the narrow width 
that decays into two Higgs bosons in the final states: $pp\to X \to HH$, $H\to b\bar{b}, WW^*$, where  both $W^{(*)}$ bosons  
decay hadronically.  Note that for such high-mass ($>1.5\,\tev$) resonance decays, more than 90\,\% of the 
events have their decay particles from the Higgs decay within a cone of $R<1.0$. 
As a result, two leading jets that have the highest and the second-highest energy with $p_T>350\gev$ and $|\eta|<2.0$ in an event 
are selected as two Higgs jet candidates. They are subsequently combined to form an $X\to HH$ resonance candidate.
The $H\to b\bar{b}$ jet identification and 
the $H\to WW^*$ jet identification are subsequently applied to reduce the background that is dominated by QCD jets 
with the criteria that both identification efficiencies are set to be 50\,\%.
The background rejection of $H\to b\bar{b}$ tagger is assumed to be the same as the center-of-mass Higgs 
tagger~\cite{ATLAS:2017juw,Aad:2020ylk} at ATLAS. 

\begin{figure}[!htb]
\begin{center}
\includegraphics[width=0.49\textwidth]{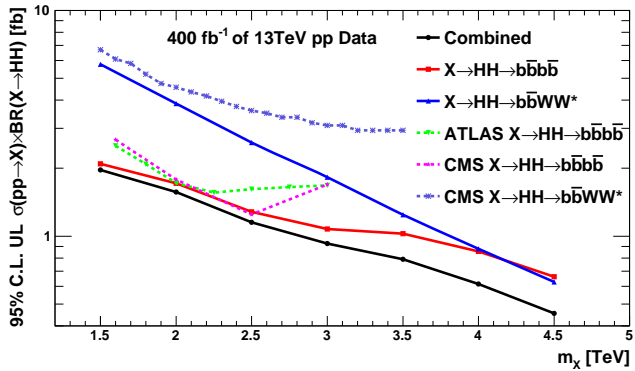}
\caption{The expected 95\% C.L. UL on the product of the production cross section of a heavy resonance $X$ and
its decaying branching faction into a Higgs boson pair, as a function of the assumed $X$ mass that is 
reconstructed in different Higgs boson decay final states.}
\label{fig:limit}
\end{center}
\end{figure}

We estimate the expected 95\,\% C.L. upper limit (UL) on the 
product of the production cross section of a heavy resonance $X$ and the branching fraction for its decay into a
Higgs boson pair. The expected limit  is  plotted as a function of the assumed $X$ mass, as shown in Fig.~\ref{fig:limit},
for $400\,\rm{fb}^{-1}$ LHC data at 13\,\tev, equivalent to the total luminosity accumulated after the incoming run III data taken
at the LHC. For heavy resonances whose masses are below 3\,\tev, the search sensitivity in the final state where both Higgs bosons decay into
a $b\bar{b}$ pair ($X\to HH\to b\bar{b}b\bar{b}$) is significantly better than the final state in which one of the Higgs bosons decays into 
two $W^{(*)}$ bosons ($X\to HH\to b\bar{b}WW^*$) because of the much larger decay branching fraction  of $H\to b\bar{b}$ 
than that of $H\to WW^*$ in the hadronic final state. However, the expected UL in the $X\to HH\to b\bar{b}WW^*$ final state
becomes comparable to the one in the $X\to HH\to b\bar{b}b\bar{b}$ channel for resonances with masses above
3.5\,\tev, due to the degradation of background rejection of $H\to b\bar{b}$ taggers for jets with very large $p_T$~\cite{ATLAS:2017juw,Aad:2020ylk}.
Besides $H\to b\bar{b}$ jets, adding $H\to WW^*$ jets as an additional experimental signature can lower 
the expected UL of  searches for $X\to HH$  by 10--50\,\%, depending on the resonance mass. 

We compare the search sensitivities in our study to the expected ones for $400\,\rm{fb}^{-1}$ data from the ATLAS and CMS experiments
by scaling down their published results~\cite{Aaboud:2018knk,Aaboud:2018zhh,Sirunyan:2018qca,Sirunyan:2019quj} according to the square root of the luminosity increase, 
as shown in Fig.~\ref{fig:EffRej}. Our estimated UL in the $X\to HH\to b\bar{b}b\bar{b}$ final state is comparable to the ones from the 
ATLAS~\cite{Aaboud:2018knk} and CMS~\cite{Sirunyan:2018qca} publications,
for resonances with masses less than 2\,\tev, but better in the higher mass region. This is because our study used a most recent 
$H\to b\bar{b}$ tagger at ATLAS~\cite{ATLAS:2017juw,Aad:2020ylk} that has a significantly better performance  than the 
Higgs taggers used in the previous ATLAS and CMS publications. 
Both the ATLAS~\cite{Aaboud:2018zhh} and CMS~\cite{Sirunyan:2019quj} experiments also carried out searches for $X\to HH\to b\bar{b}WW^*$ in the
final state where one $W^{(*)}$ boson decays hadronically and the other $W^{(*)}$ boson decays leptonically. 
Depending on the resonance mass,  the expected UL from the CMS (ATLAS) publication is approximately
1.2--2 ($>10$) times the expected UL in our study, where both hadronically decaying $W^{(*)}$ bosons are reconstructed as a single jet. 

Note that the above comparisons do not have a strong dependence on the theoretical models used to generate heavy resonances as long as the
resonances have much narrower widths than the detector resolution of the reconstructed $X\to HH$ candidates, which
is typically a few hundred \gev, and increases when the masses of the resonances become larger.

\section{conclusion}
In this paper we study the reconstruction and identification of $H\to WW^*$ with high transverse momentum,
 where both $W^{(*)}$ bosons decay hadronically. We show that the boosted $H\to WW^*$ can be effectively reconstructed as a single jet and 
 identified using jet substructures in the center-of-mass frame of the jet.  
Such a  reconstruction and identification approach can  discriminate the boosted 
$H\to WW^{*}$ in the full hadronic final state from QCD jets.  Our result will significantly improve 
experimental sensitivities of searches for potential NP beyond the SM in final states containing highly boosted Higgs bosons.
 The technique we proposed can be also directly applied to boosted $H\to ZZ^{*}$ in the full hadronic final state.

\section{Acknowledgments}

This work is supported by the Office of Science of the U.S. Department of
Energy under Contract No. DE-FG02-13ER42027.

\end{document}